\documentclass[twocolumn,showpacs,preprintnumbers,amsmath,amssymb,superscriptaddress]{revtex4}

\usepackage{longtable}
\usepackage{graphicx}
\usepackage{dcolumn}
\usepackage{bm}

\begin{document}

\title{Measuring Double-Electron Capture with Liquid Xenon Experiments}

\newcommand{\usd}{Department of Physics, The University of South Dakota, Vermillion, South Dakota 57069}
\newcommand{\yru}{College of Physics, Yangtz River University, Jingzhou 434023, China}
\newcommand{\tgu}{College of Sciences, China Three Gorges University, Yichang 443002, China}
\newcommand{\dmm}{Dongming.Mei@usd.edu}

\author{     D.-M.~Mei          }\altaffiliation[Corresponding Author: ]{\dmm}	\affiliation{ \usd } 	\affiliation{ \yru }
\author{I. Marshall} \affiliation{ \usd} 
\author{W.-Z. Wei}\affiliation{ \usd}
\author{    C. Zhang           }\affiliation{ \usd} \affiliation{\tgu} 			

\begin{abstract}
We investigate the possibilities of observing the decay mode for $^{124}$Xe in which two electrons are captured, two neutrinos are emitted, 
and the final daughter nucleus is in its ground state, using dark matter 
experiments with
liquid xenon. The first upper limit of the decay half-life is calculated to be 1.66$\times$10$^{21}$ years at a 90\% confidence level (C.L.)  
obtained with the published background data from the XENON100 experiment. Employing a known
background model from the Large Underground Xenon (LUX) experiment, we predict that the detection of double-electron capture
 of $^{124}$Xe to the ground state of $^{124}$Te with LUX will have approximately 115 events, assuming a half-life of 2.9 $\times$ 10$^{21}$ years. 
We conclude that measuring 
$^{124}$Xe 2$\nu$ double-electron capture to the ground state of $^{124}$Te 
can be performed more precisely with the proposed LUX-Zeplin (LZ) experiment. 
  
\end{abstract}

\maketitle
\section{Introduction}
The decay mode of an atomic nucleus in which two of the orbital electrons are captured by two protons and two neutrinos are emitted 
in the process is called two neutrinos double-electron capture (2$\nu$DEC)~\cite{mey,war,yak}. 
Equation~(\ref{eq:ecec2v}) shows the decay process.
\begin{equation}
2e^{-} + (Z,A) \rightarrow (Z-2, A) + 2\nu_{e},
\label{eq:ecec2v}
\end{equation}
where Z is the atomic number, and A is the atomic mass number for a given nucleus. The positive results were reported by a geochemical experiment~\cite{apm} 
for $^{130}$Ba with a half-life of  (2.2$\pm$0.5)$\times$10$^{21}$ years and a noble gas experiment~\cite{ymg} for $^{78}$Kr with
a half-life of (9.2$^{+5.5}_{-2.6}$(stat)$\pm$1.3(syst))$\times$10$^{21}$ years. The 2$\nu$DEC process is allowed by 
the Standard Model of particle physics and no conservation laws (including lepton number conservation) are violated. 

If two electrons are captured by two protons in the nucleus, and neutrinos are {\it not} emitted, the process is called 
neutrinoless double-electron capture (0$\nu$DEC)~\cite{zsu} in which the lepton number is {\it not} conserved, 
 and the neutrino is its own antiparticle, a Majorana particle. If observed, this mode of decay described in equation~(\ref{eq:0vecec})
would require new particle physics 
beyond the Standard Model. 
\begin{equation}
2e^{-} + (Z,A) \rightarrow (Z-2, A).
\label{eq:0vecec}
\end{equation}

\begin{figure}[htb!!!]
\includegraphics[angle=0,width=6.cm] {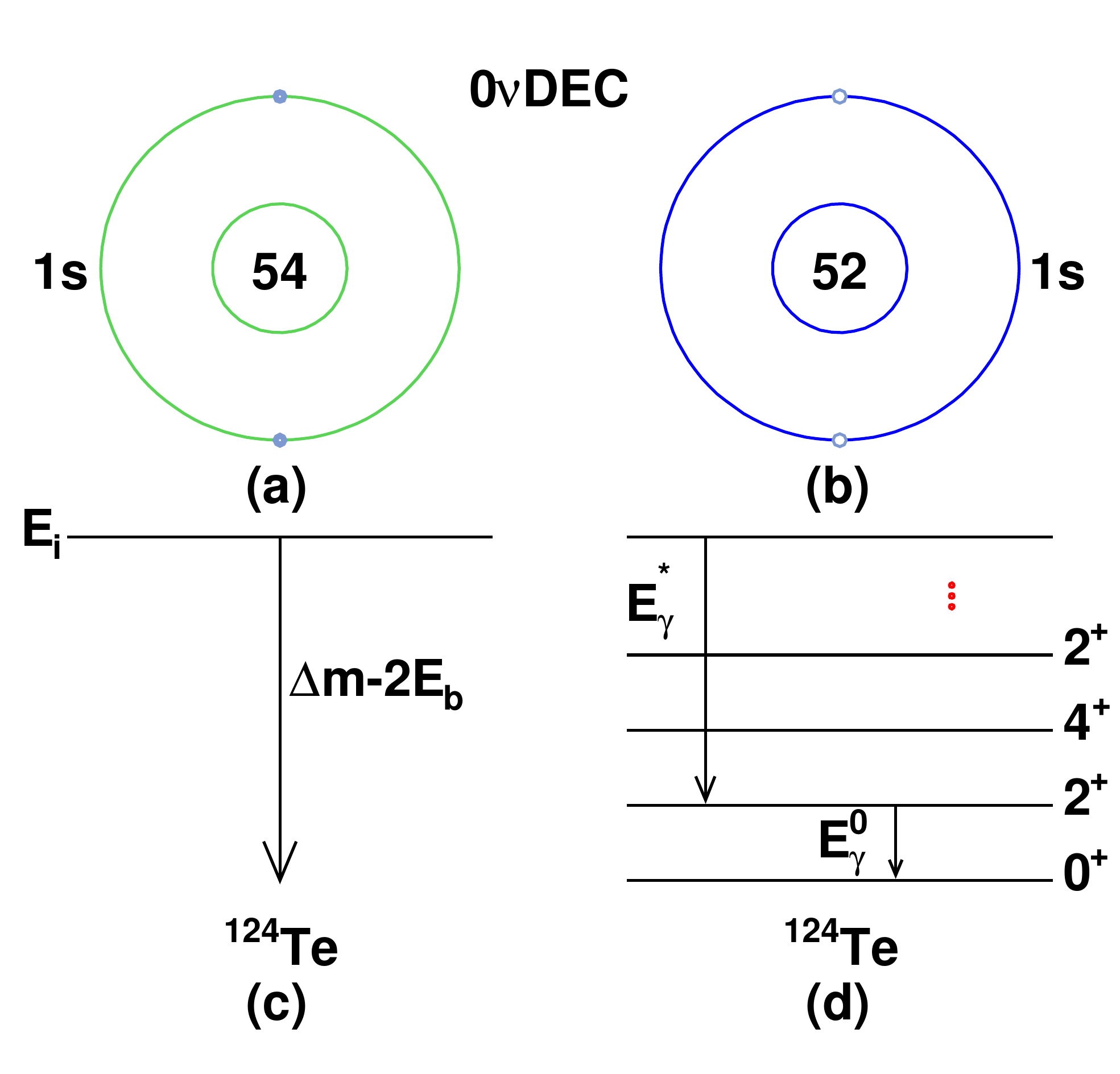}
\caption{\small{(Color online) The schematic diagram for the $^{124}$Xe 0$\nu$DEC process. 
(a) $^{124}$Xe has 54 electrons (two electrons in the 1s shell) and 54 protons. (b) $^{124}$Te has 52 electrons (two holes in the 1s shell) 
and 52 protons in an excited state. Both can emit electromagnetic radiation through de-excitation. The atomic de-excitation is shown in (c). 
(d) shows the nuclear de-excitation. E$^{*}_{\gamma}$ is the $\gamma$-ray which will signal double electron capture originating 
from a state with energy $\Delta$$m-2b$ which has tiny admixtures of 0$^{+}$, 2$^{+}$, 4$^{+}$, etc.}}
\label{fig:mech}
\end{figure}

The experimental study of this process is very challenging due to its extremely long lifetime. This is because the decay process 
is expected
to be accompanied by an internal Bremsstrahlung gamma quantum and the final nucleus is in an excited state, which strongly 
suppresses the allowed decay
phase space~\cite{verg, doim}.  
In contrast to neutrinoless double-beta decay, a rare process used as a powerful tool to test neutrino properties and 
lepton number violation with several on-going experiments~\cite{gerda, cobra, nemo, cuore, majorana, kam, exo}, 
neutrinoless double-electron capture 
appears to be extremely slow as pointed out by Vergados, J.D.~\cite{verg} and discussed in detail by Doi, M. and Kotani, 
T.~\cite{doim}. However, a possible resonant 0$\nu$DEC process in which the close degeneracy of the initial and final (excited) atomic states 
can enhance the decay rate by a factor as large as 10$^{6}$~\cite{bjd}, which might occur, has been studied by
many authors~\cite{bjd, jsu, sju, sel, green, bas, bsa, sab, mfk, suj}. The 0$\nu$DEC might be realized as a resonant 
decay~\cite{bjd, jsu, sel, green, bas, bsa, sab, mfk} 
or as a radiative process with or without a resonance condition~\cite{suj}.
Figure~\ref{fig:mech} shows an example of 0$\nu$DEC with $^{124}$Xe. 

Nonetheless, 
the 2$\nu$DEC process is a standard nuclear process and it should be detected experimentally by measuring its half-life, 
as expressed below:
\begin{equation}
(T_{\frac{1}{2},2\nu})^{-1} = \frac{a_{2\nu}F_{2\nu}|M_{2\nu}|^{2}}{ln(2)},
\label{eq:halflife}
\end{equation}
where $a_{2\nu}$$\sim$2$\times$10$^{-22}$y$^{-1}$ is the dimensional factor, F$_{2\nu}$ is the phase-space factor (proportional 
to Q$^{5}$), and M$_{2\nu}$ is the nuclear matrix element (NME). 

The measurement of the two X-ray energies and the half-life of the 2$\nu$DEC decay (T$_{\frac{1}{2},2\nu}$) to the ground state is of great 
interest to nuclear physics. Of particular interest is,
how the double $K$ vacancy refills after the capture of two electrons by two protons in the nucleus occurred. Measuring the total energy 
from X-rays can shed 
some light on the precise mechanism of this atomic decay process. Moreover, if the mass difference between the initial and final states 
is greater than twice the mass of electron (1.022 MeV), the reaction Q value is enough to initiate another mode of decay, which would be 
electron capture and positron emission. This decay mode occurs in competition with double-electron capture and their branching ratio 
depends 
on nuclear properties, which is of great interest. Furthermore, 
when the mass difference is greater than four electron masses (2.044 MeV), the third mode - double-positron decay - can occur as well. 
However, only 6 naturally occurring nuclides can decay via these three modes simultaneously~\cite{frek}. $^{124}$Xe, discussed below, is one of them. 
Therefore, measuring the decay modes of $^{124}$Xe has particular meaning in nuclear physics. 
In addition, the
model predictions for 0$\nu$DEC half-life require the evaluation of nuclear matrix elements. These calculations are complicated and 
have large uncertainties. 
They are different from those required for 2$\nu$DEC. However, within the same model framework,
 some constraints
 on the 0$\nu$DEC NME0$\nu$ can be derived using knowledge of the 2$\nu$DEC NME2$\nu$~\cite{green, howard}. Also, the NME2$\nu$,
 which is extracted from the measurement 
of the half-life of 2$\nu$DEC, can be directly compared with the NME2$\nu$ from predictions~\cite{verg}. 
A good agreement would indicate that 
the reaction mechanisms and the nuclear structure aspects that are involved in 2$\nu$DEC are well understood.

2$\nu$DEC has large Q-values, but the decay to ground state in the final nucleus releases only X rays and Auger electrons, 
making its detection difficult. At their energy range ($\sim$1 to $\sim$100 keV), the background is usually high. Thus, 
the experimental detection of double electron capture with 2$\nu$ emission is more difficult than 2$\nu$ double beta decay, which
has been observed for a variety of nuclei~\cite{elliott, kdc, gerda1, nemo, supernemo1, supernemo2, supernemo3, supernemo4, exo1}.  
 Nevertheless, experiments
directly searching for dark matter require ultra-low background events in the low energy region (down to $\sim$1 to $\sim$100 keV). 
 This lays the foundation to experimentally measure 2$\nu$DEC process for  the first time. In this paper, we discuss
the detection of 2$\nu$DEC with $^{124}$Xe in the dark matter experiments with natural xenon as targets, such as 
XENON100~\cite{xenon100}, LUX~\cite{lux}, and LUX-Zeplin (LZ)~\cite{lz}.  

Natural xenon possesses $^{124}$Xe at an abundance of 0.1\%~\cite{gru,jrd}. The process for 2$\nu$DEC of $^{124}$Xe is:
\begin{equation}
^{124}Xe + 2e^{-} \rightarrow ^{124}Te + 2\nu_{e}.
\label{eq:xe124}
\end{equation}

The reaction Q value is 2864 keV. For the ground state of $^{124}$Xe to the ground state of $^{124}$Te, the detectable X rays are 31.8 keV 
from $^{124}$Te, for a one-step process in which the two $K$-shell electrons are captured simultaneously by two protons in the nucleus.
 The nuclear recoil
energy of $^{124}$Xe allocated in the decay process is on the order of $\sim$30 eV, which is negligible. 
The predicted half-life for 2$\nu$DEC
is 2.9$\times$10$^{21}$ years~\cite{bara} for a ground state to ground state process.

Since the reaction Q value in equation~(\ref{eq:xe124}) is 2864 keV, the two other decay modes - electron capture with positron 
 (2$\nu\beta^{+}$EC) emission and double positron decay (2$\nu\beta^{+}\beta^{+}$) can simultaneously occur with double electron capture.
The available energies are shown below: 
\begin{eqnarray}
\left\{ \begin{array}{c}
Q_{DEC} = M(A,Z) - M(A, Z-2),\\
Q_{\beta^{+}EC} = M(A,Z) - M(A, Z-2) - 2mc^{2},\\
Q_{\beta^{+}\beta^{+}} = M(A,Z) - M(A,Z-2) - 4mc^{2}.
\end{array} \right.
\label{eq:qvaule}
\end{eqnarray}
However, 
the 2$\nu$DEC rate is much faster
than 2$\nu\beta^{+}$EC and  2$\nu\beta^{+}\beta^{+}$ as discussed in Refs~\cite{jsu, bara}.

It is worth mentioning that $^{126}$Xe has also a natural abundance of 0.09\% and 
can only undergo a 2$\nu$DEC or a 0$\nu$DEC decay, $^{126}$Xe$\rightarrow$$^{126}$Te, with $^{126}$Te at its ground state 
and the total decay Q value of 896
keV. Because this decay Q value, 896 keV, is a factor of 3.2 smaller than the Q value, 2864 keV, from $^{124}$Xe decays, the
$^{126}$Xe 2$\nu$DEC is much slower than $^{124}$Xe 2$\nu$DEC decay. Therefore, we will not discuss $^{126}$Xe 2$\nu$DEC in this paper.
 
\section{The first upper limit of half-life from xenon-124}
The average upper limit in an experiment with background can be obtained using the unified approach proposed by 
G. Feldman and R. Cousins~\cite{gfrc}. For a detector with $^{124}$Xe target, the upper limit of the half-life can be derived  
using the following equations~\cite{yzde}:
\begin{eqnarray}
T_{1/2} (0^{+} \rightarrow g.s.) \geq \frac{ln(2) \cdot f_{k} \cdot \epsilon \cdot a \cdot \frac{M \cdot N_{A}}{A} \cdot \Delta T} {\mu_{up}},\\
\mu_{up} \cong \alpha \cdot \sqrt{B},\\
B = b \cdot \Delta T \cdot \Delta E, 
\label{eq:sens}
\end{eqnarray}
where $f_{k}$ is the fraction of 2$K$ captures accompanied by the emission of two $K$ X rays, 
$\epsilon$ is the efficiency of the detection at a full energy peak, $a$ is the isotopic abundance of $^{124}$Xe, and M is the 
total mass of 
the target. N$_{A}$ is Avogadro constant, A is the atomic mass number of $^{124}$Xe, $\Delta T$ is the live time of measurements in days,
$\alpha$ is a constant that equals to 1.64 at 90\% confidence level (C.L.),  $b$ is 
the background rate per unit energy, and $\Delta E$ is the energy window around the peak position.
\subsection{Results and analysis from XENON100} 
\begin{figure}[htb!!!]
\includegraphics[angle=0,width=8.cm] {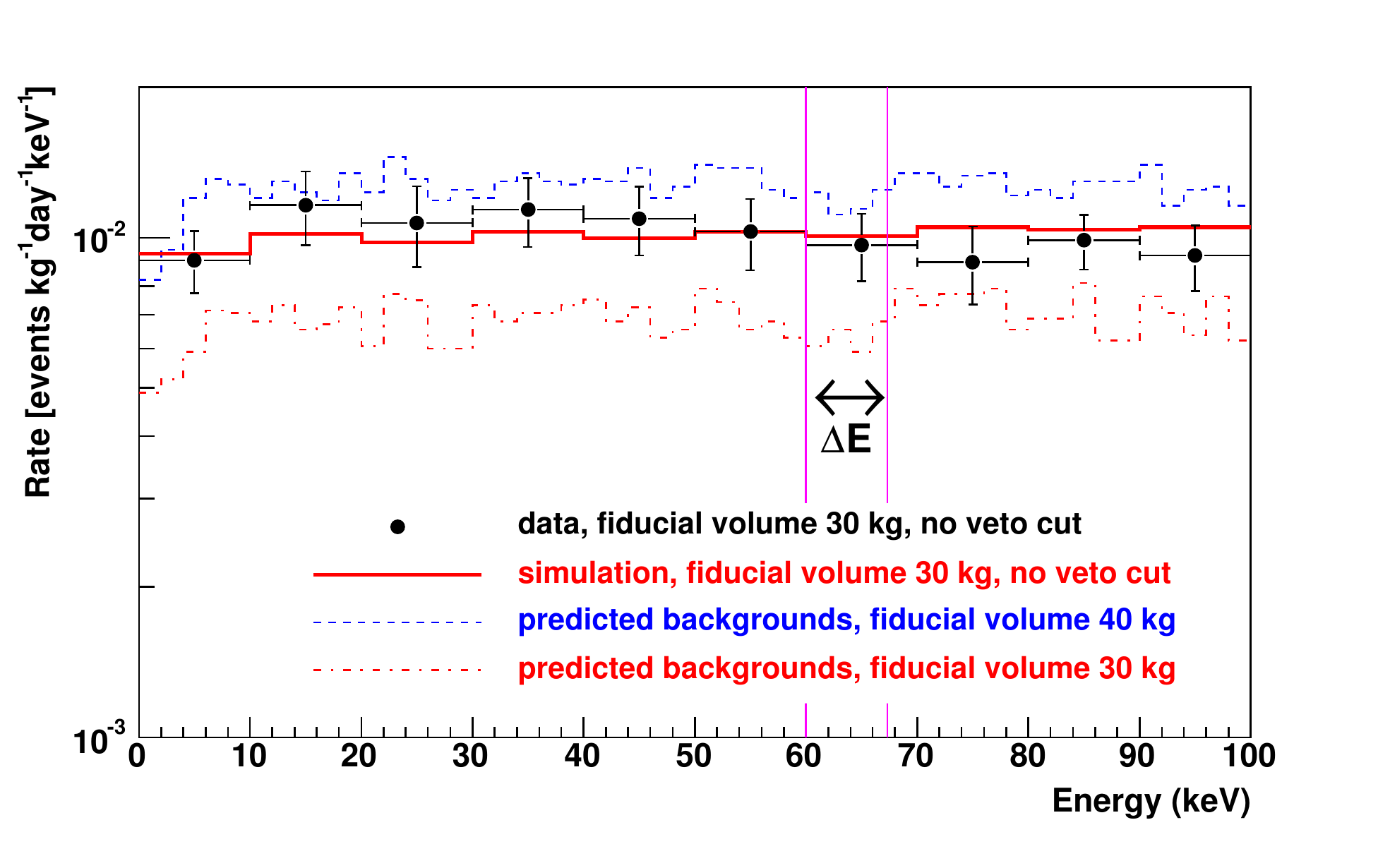}
\caption{\small{ (Color online) Shown is a digitized background spectrum from the published XENON100 data~\cite{eap}.
The region of interest, 63.6$\pm$$\Delta$E keV, is labelled.   }}
\label{fig:xenon100analysis}
\end{figure}

The XENON100 dark matter experiment reported their dark matter analysis with a 34 kg active target of liquid xenon~\cite{eapr}.
The electromagnetic background events in the region of interest was reported as 5.3$\times$10$^{-3}$ events/(kg\hspace{0.03in}day\hspace{0.03in}keV).
We analyzed the XENON100 electromagnetic background data with a digitized spectrum, as shown in Figure~\ref{fig:xenon100analysis}, 
from the published background spectra (Figure 3 and Figure 12 in Ref.~\cite{eap}) between 0 to 100 keV. A peak-searching algorithm, 
Wavelet Transform~\cite{wee}, was applied in searching for peaks and no peak was found in the region of interest as shown in 
Figure~\ref{fig:xenon100analysis}. 

Consequently, an average background rate, 5.3$\times$10$^{-3}$ events/(kg\hspace{0.03in}day\hspace{0.03in}keV), was used 
in the calculation of the background index, events/(keV\hspace{0.03in}day), in Table~\ref{tab:xenon100}. 
The width of the region of interest, $\Delta$E = 7.94 keV, is determined using $\alpha$$\times$$\sigma$, 
where $\alpha$ equals 1.64 at 90\% C.L. and $\frac{\sigma}{E}$ =
0.009+0.485/$\sqrt{E(keV)}$~\cite{eap} is
energy resolution, $E$ is the sum of the expected two X-rays (2$\times$31.8 keV) from $^{124}$Te. In addition, Gavrilyuk et al. reported 
that the energy released in the refilling of a double K-vacancy is not equal to the sum of 
two single vacancies~\cite{ymg}. Therefore, a possible reduction of the total energy release (2$\times$31.8 keV) was taken into account, using an energy window of 7.94 keV, in our analysis. 
This possible energy reduction is from fluorescence yield, which might not be detectable in liquid xenon, induced by the emission of Auger electrons.
Because the position resolution is less than
3 mm~\cite{eap}, the detection efficiency for two X-rays can be 90\% since the the mean free path of X-rays with energy of 31.8 keV 
in liquid xenon is about 0.5 mm. The determined analysis parameters are summarized in Table~\ref{tab:xenon100}.

\begin{figure}[htb!!!]
\includegraphics[angle=0,width=8.cm] {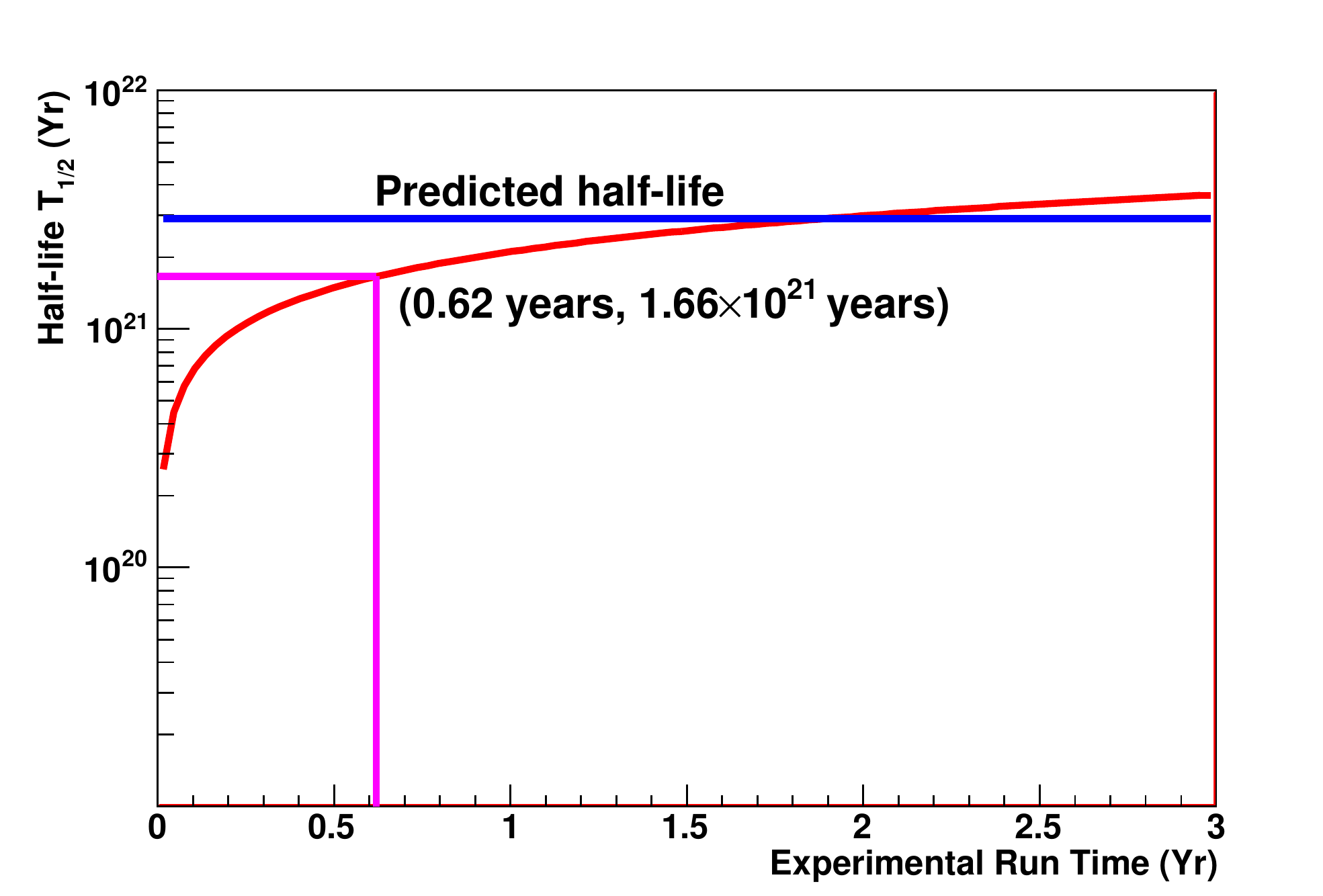}
\caption{\small{(Color online) Above is the half-life limit of the $^{124}$Xe 2$\nu$DEC process to the ground state of $^{124}$Te. }}
\label{fig:plot1}
\end{figure}

Using
equations~(6-\ref{eq:sens}) and the values given in Table~\ref{tab:xenon100}, the half-life limit of $^{124}$Xe 2$\nu$DEC to 
its ground state
is determined to be 1.66$\times$10$^{21}$ years with a 90\% C.L. (1.64$\sigma$), as shown in Figure~\ref{fig:plot1}.

\begin{table}[htb!!!]
\caption{The experimental parameters and values. }
\label{tab:xenon100}
\begin{tabular}{|c|c|}
\hline \hline
Mass of liquid xenon, kg & 34\\ 
Isotope abundance, \% & 0.1\\
Live time, days & 225\\
Background index, events/(keV\hspace{0.03in}day) & 0.18\\
K-shell fluorescence yields ($\omega_{k}$) &0.875~\cite{wal}\\
$f_{k}$ = $\omega_{k}^{2}$ &0.766\\
Efficiency at 63.6 keV & 0.9\\
Energy resolution ($\frac{\sigma}{E})$ at 63.6 keV, \%& 7.0\\
The region of interest $\Delta E$, keV & 7.94\\ 
\hline
Reaction Q value, keV & 2864\\
\hline \hline
\end{tabular}
\end{table}

\subsection{Predicted results from LUX-like and LZ-like experiments}
\begin{figure}[htb!!!]
\includegraphics[angle=0,width=8.cm] {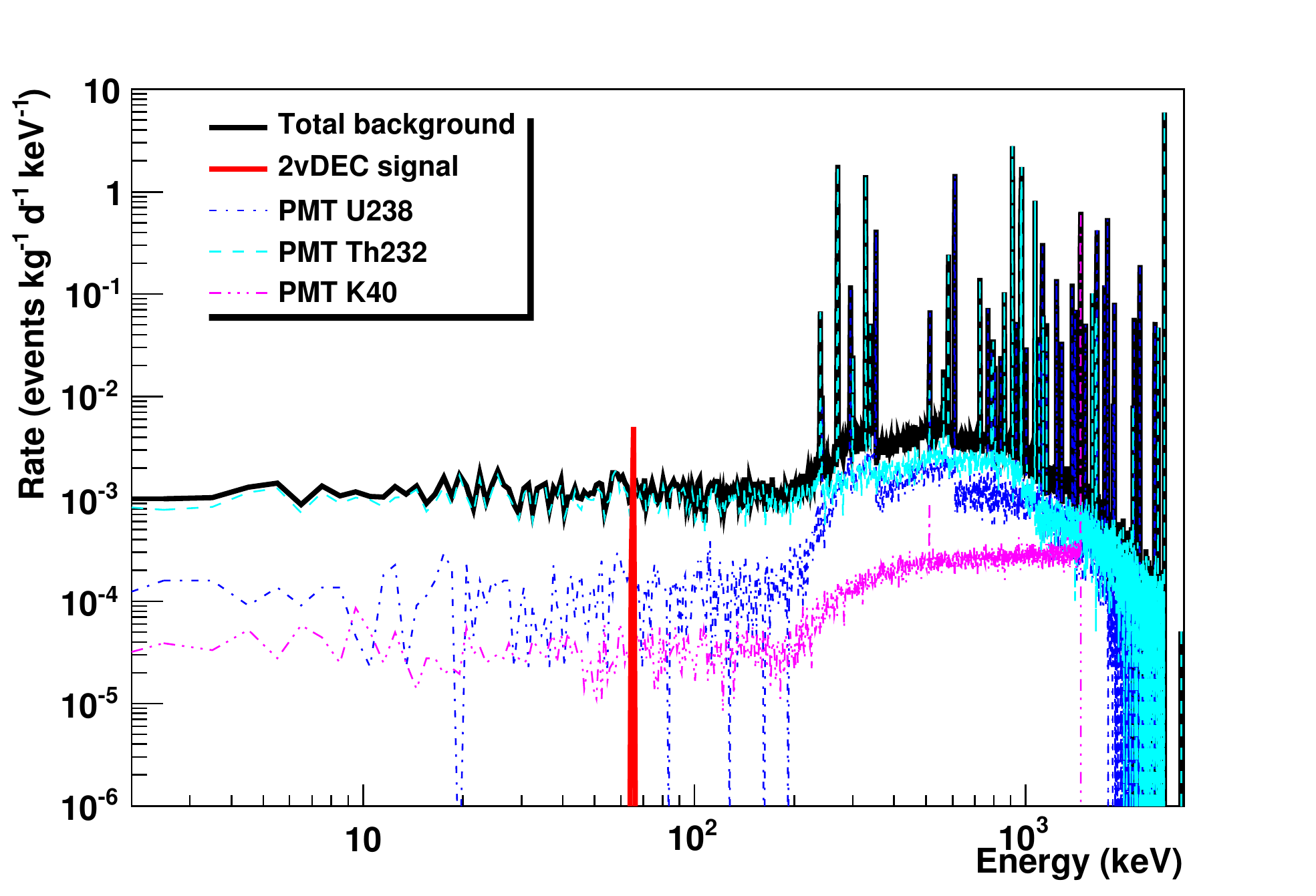}
\caption{\small{(Color online) Simulated signal events for $^{124}$Xe 2$\nu$DEC to the ground state of $^{124}$Te 
in the LUX detector with a known background model. }}
\label{fig:plot2}
\end{figure}

The LUX dark matter experiment has been constructed at Sanford Underground Research Facility (SURF)~\cite{lux} 
and is currently taking data.
The LUX detector contains
360 kg of xenon with an assumed fiducial volume of 100 kg. 
We calculated the detectable events, for $^{124}$Xe to the ground state of $^{124}$Te,  to be approximately 115 per year, 
assuming the predicted half-life is 2.9$\times$10$^{21}$ years.
From a  background model published with a  Monte Carlo 
simulation~\cite{luxsim} for the LUX detector, we know the dominant background is from the PMT sphere, which has radioactivity contents shown in
Table~\ref{tab:lux}. 
\begin{table}[htb!!!]
\caption{Radioactivity level of the LUX 8778 PMT~\cite{lux}. Units are in mBq/PMT.}
\label{tab:lux}
\begin{tabular}{|c|c|c|c|}
\hline \hline
$^{238}$U & $^{232}$Th & $^{60}$Co & $^{40}$K\\ 
9.8$\pm$0.7 & 2.3$\pm$0.5& 2.2$\pm$0.4 & 65$\pm$2\\
\hline \hline
\end{tabular}
\end{table}
Using the radioactivity levels in Table~\ref{tab:lux}, a simple Monte Carlo simulation was performed to predict the signal events from $^{124}$Xe DEC process together with several significant sources of background from PMTs, Figure~\ref{fig:plot2} shows the predicted results.

\begin{figure}[htb!!!]
\includegraphics[angle=0,width=8.cm] {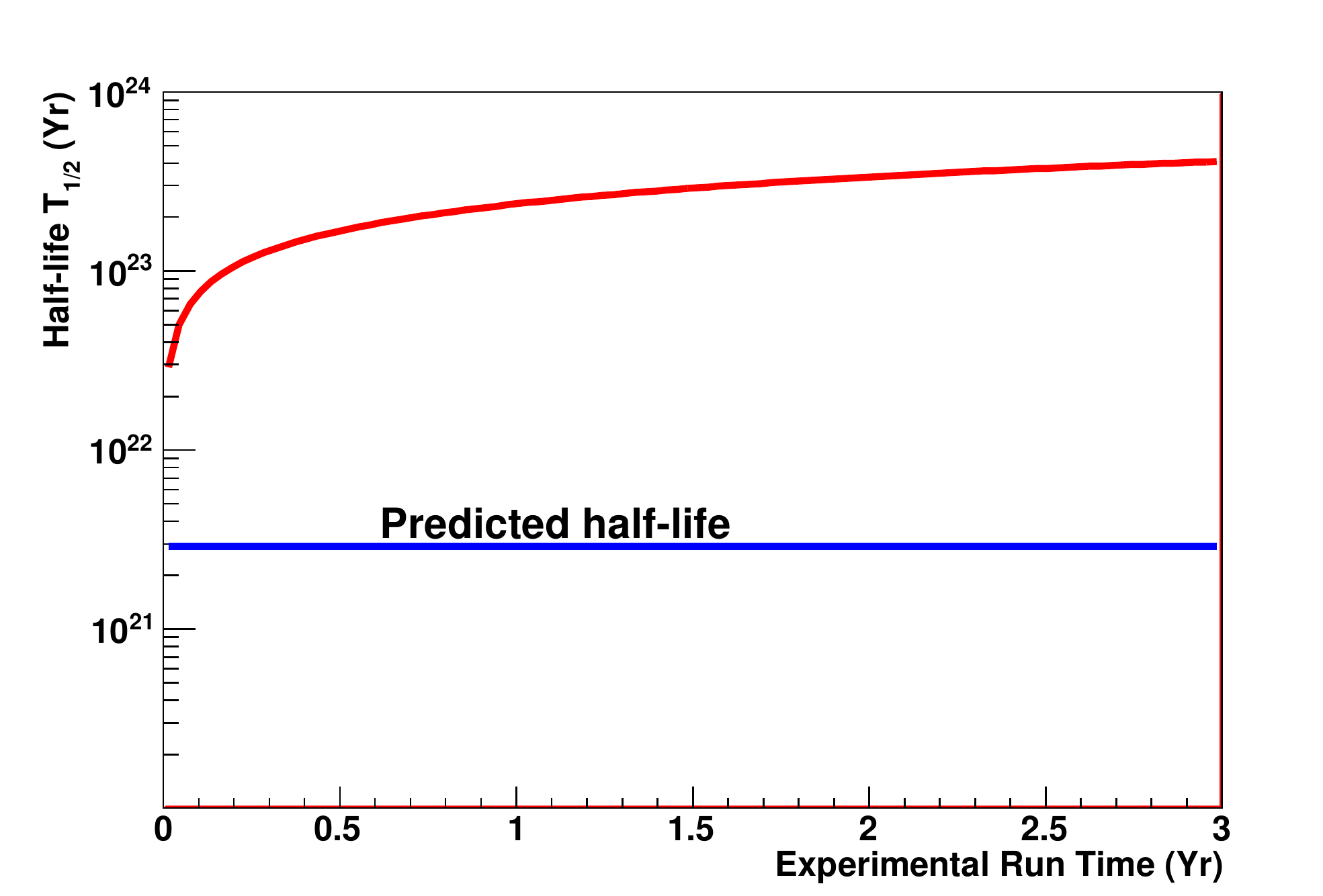}
\caption{\small{(Color online) The sensitivity of the half-life limit for $^{124}$Xe 2$\nu$DEC to the ground state of $^{124}$Te, 
utilizing the LZ experiment. }}
\label{fig:plot3}
\end{figure}

The proposed LUX-Zeplin (LZ) experiment will contain 7 tons of liquid xenon~\cite{lz}.  Although the mass of xenon in LZ is only
20 times greater than in LUX~\cite{lux}, the expected sensitivity of LZ will exceed that of LUX by over two orders of magnitude.  
The additional sensitivity, which is greater than a simple scaling of xenon mass, is due primarily to improved background suppression. 
This, in turn, enables a longer running time for the LZ experiment and allows a larger effective fiducial mass fraction 
after the projected 
analysis cuts. Figure~\ref{fig:plot3} shows a sensitivity plot for measuring 2$\nu$DEC using the LZ detector, assuming a background rate
of 1.8$\times$10$^{-4}$/(kg\hspace{0.03in}keV\hspace{0.03in}day) at the region of interest. 

\section{Conclusion}
We have derived the first upper limit of the two neutrino double-electron capture process for $^{124}$Xe to the ground state of $^{124}$Te
 using published XENON100 experimental 
data. 
The obtained upper limit of 1.66$\times$10$^{21}$ years was compared to the predicted half-life of 2.9$\times$10$^{21}$ years, which can
 be 
measurable from the XENON100 experiment in three more years. Utilizing the published LUX background model, we predicted approximately
 115 events per year in 
the LUX detector, assuming a half-life of 2.9$\times$10$^{21}$ years. These 115 events are measurable with the
LUX background model. By comparing our predicted events from the LUX detector to the more sensitive and larger LZ detector,
 we should be able to confidently measure this process. 

\section{Acknowledgement}
The authors wish to thank Christina Keller for careful reading of this manuscript. 
In particular, the authors 
would like to thank D$^{'}$Ann Barker for useful discussion and the members of the LZ collaboration for their
many invaluable suggestions in preparing this paper.  
This work was supported in part by NSF PHY-0758120, NSF PHY-0919278, NSF PHY-1242640, DOE grant DE-FG02-10ER46709,
 the Office of Research at the University 
of South Dakota and a 2010 research center supported by the state of South Dakota. 

%
%

\begin{thebibliography}{99}
\bibitem{mey} W.E. Meyerhof et al., Phys. Rev. A 32 (1985) 3291.
\bibitem{war} Warczak, A. et al., Nucl. Instar. Meth. B 98 (1995) 303.
\bibitem{yak} V.L. Yakhontov, M.Y. Amusia, Phys. Rev. A 55 (1997) 1952.
\bibitem{apm} A. P. Meshik, C. M. Hohenberg, O. V. Pravdivtseva, Y. S. Kapusta,  Phys. Rev. C 64 (2001) 035205.
\bibitem{ymg} Yu. M. Gavrilyuk, A. M. Gangapshev, V. V. Kazalov, and V. V. Kuzminov, Phys. Rev. C 87 (2013) 035501.
\bibitem{zsu} Z. Sujkowski and S. Wycech, Phys. Rev. C 70 (2004) 052501 (R).
\bibitem{verg} Vergados, J.D., Nucl. Phys. B218 (1983) 109.
\bibitem{doim} Doi, M., Kotani, T., Progr. Theor. Phys. 89 (1993) 139.
\bibitem{bjd} Bernabeu J, De Rujula A and Jarlskog C (1983) Nucl. Phys. B 223 15–28.
\bibitem{jsu} Jouni Suhonen, Journal of Physics: Conference Series 375 (2012) 042026.
\bibitem{sju} Jouni Suhonen, J. Phys. G: Nucl. Part. Phys. 40 (2013) 075102.
\bibitem{sel} S. Eliseev et al., Phys. Rev. Lett. 106 (2011) 052504.
\bibitem{green} K. L. Green et al., Phys. Rev. C 80 (2009) 032502.
\bibitem{bas} A. S. Barabash et al., Nucl.Phys.Proc.Suppl. 229-232 (2012) 474.
\bibitem{bsa} A. S. Barabash et al., Phys.Rev. C83 (2011) 045503.
\bibitem{sab} A. S. Barabash et al., J.Phys.Conf.Ser. 120 (2008) 052057.
\bibitem{mfk} M. F. Kidd, J. H. Esterline, and W. Tornow, Phys.Rev. C78 (2008) 035504.
\bibitem{suj} Sujkowski Z and Wycech S, Phys. Rev. C 70 (2004) 052501.
\bibitem{gerda} A. Bettini for the GERDA Collaboration, Nucl. Phys. B 168 (2007) 67.
\bibitem{cobra} Zuber K., AIP Conference Proceedings 1180 (2009) 145.
\bibitem{nemo} Chauveau E., AIP Conference Proceedings 1180 (2009) 26.
\bibitem{cuore} Arnaboldi C. et al., Phys. Lett. B584 (2004) 212-213.
\bibitem{majorana} Detwiler, J.; et al. (2011). "The M {\sc AJORANA} D{\sc EMONSTRATOR}". arXiv:1109.6913.
\bibitem{kam} A. Gando et al., arXiV:1211.3863v1.
\bibitem{exo} M. Auger et al., Phys.Rev.Lett. 109 (2012) 032505.
\bibitem{frek} D. Frekers arXiv:help-ex/0506002v2.
\bibitem{howard} Howard M. Georgi et al., Nucl. Phys. B193 (1981) 297-316.
\bibitem{elliott} S. R. Elliott, A. A. Hahn, M. K. Moe, Phys.Rev.Lett. 59 (1987) 2020-2023.
\bibitem{kdc} Klapdor-Kleingrothaus HV et al (Heidelberg-Moscow Collaboration) 2001 Eur. Phys. J. A 12 147.
\bibitem{gerda1} M. Agostini et al., J. Phys. G: Nucl. Part. Phys; J. Phys. G: Nucl. Part. Phys. 40 (2013) 035110.
\bibitem{supernemo1} R. Arnold, et al.: Phys. Rev. Lett. 95 (2005) 182302.
\bibitem{supernemo2} Argyriades J et al., Phys. Rev. C 80 (2009) 032501R.
\bibitem{supernemo3} Argyriades J et al., Nucl. Phys. A 847 (2010) 168.
\bibitem{supernemo4} Arnold R et al., Phys. Rev. Lett. 107 (2011) 062504.
\bibitem{exo1} Ackerman, N. et al., Phys. Rev. Lett. 107 (2011) 212501. 
\bibitem{xenon100} E. Aprile et al., arxiV:1107.2155V2.
\bibitem{lux} D. S. Akerib et al., LUX Collaboration, Nucl. Instrum. Meth. A 704 (2013) 111-126.
\bibitem{lz} Private communication, The LZ collaboration.
\bibitem{gru} G. Audi et al., Nuclear Physics A729 (2003) 3-128.
\bibitem{jrd} J. R. de Laeter et al., Pure and Applied Chemistry 78 (11): 2051–2066. doi:10.1351/pac200678112051.
\bibitem{bara} A. S. Barabash et al. Phys. Lett. B223, 273 (1989).
\bibitem{gfrc} Gary J. Feldman and Robert D. Cousins, Phys. Rev. D57 (1998) 3873.
\bibitem{yzde} Y. Zdesenko, Review of Mondern Physics, Vol. 74 (2002) 663-684.
\bibitem{eapr} E. Aprile, M. Alfonsi, K. Arisaka et al. (XENON100 Collaboration), Phys. Rev. Lett. 109 (2012) 181301. 
\bibitem{eap}  E. Aprile, M. Alfonsi, K. Arisaka et al. (XENON100 Collaboration), Phys. Rev. D83 (2012) 082001.
\bibitem{wal} W. Bambynek et al., Reviews of Modern Physics 44 (1972) 716. 
\bibitem{wee}  Andrew Wee, David B. Grayden, Yonggang Zhu, et al., Electrophoresis 29 (2008) 4215 - 4225.
\bibitem{luxsim} D. S. Akerib et al., Nucl. Instr. and Meth. A 675 (2012) 63-77.


\end{thebibliography}

\end{document}